\documentclass[12pt,titlepage]{article}
\usepackage{amsfonts}
\usepackage{amsmath}
\usepackage{amssymb}
\usepackage{times}
\usepackage[doublespacing]{setspace}
\usepackage{mathrsfs}
\usepackage{geometry}
\usepackage{rotating}

\input{tcilatex}
\renewcommand{\mathcal}{\mathscr}
\begin{document}

\title{Warped Functional Analysis of Variance}
\author{Daniel Gervini \\
\emph{Department of Mathematical Sciences}\\
\emph{University of Wisconsin--Milwaukee}\\
\emph{PO Box 413, Milwaukee, WI 53201} \and and Patrick A. Carter \\
\emph{School of Biological Sciences}\\
\emph{Washington State University}\\
\emph{PO Box 644236, Pullman, WA 99164 }}
\maketitle

\begin{abstract}
This article presents an Analysis of Variance model for functional data that
explicitly incorporates phase variability through a time-warping component,
allowing for a unified approach to estimation and inference in presence of
amplitude and time variability. The focus is on single-random-factor models
but the approach can be easily generalized to more complex ANOVA models. The
behavior of the estimators is studied by simulation, and an application to
the analysis of growth curves of flour beetles is presented. Although the
model assumes a smooth latent process behind the observed trajectories,
smootheness of the observed data is not required; the method can be applied
to the sparsely observed data that is often encountered in longitudinal
studies.

\end{abstract}

\section{\label{sec:Introduction}Introduction}

The main motivation for the present paper is the study of functional traits
in evolutionary biology and quantitative genetics. Evolutionary biology
investigates the change of physical traits (phenotypes) across generations.
Some traits are univariate or multivariate, but others are functional, like
growth curves or thermal performance curves (Kirkpatrick and Heckman, 1989;
Heckman, 2003; Kingsolver et al., 2002; Meyer and Kirkpatrick, 2005; Ragland
and Carter, 2004). Understanding the modes of variability of these curves is
important in order to understand the biological processes behind the trait,
and in particular the genetic aspects of it.

Consider for example the flour-beetle growth curves shown in Figure \ref%
{fig:curves}(a) (see Irwin and Carter, 2013, for details about these data).
They are mass measurements of larvae from hatching to pupation. The dataset
consists of 122 half-siblings sired by 29 fathers and different mothers. A
distinct characteristic of these curves is an inflection point around day
15; this is the time when larvae stop eating and begin searching for a place
to pupate. This process is triggered by hormonal mechanisms whose timing
varies from individual to individual; determining what proportion of the
time variability can be attributed to genetic factors and what proportion
can be attributed to environmental factors is important for understanding
the evolution of development and growth. Similarly, in the study of thermal
performance curves (which are functions of temperature, not time), the
optimal temperature varies from individual to individual and characterizing
the sources of this variability is important for understanding thermal
adaptations (Huey and Kingsolver, 1989; Izem and Kingsolver, 2005).

We can see, then, that functional samples usually present two types of
variability: what we can denominate \textquotedblleft
horizontal\textquotedblright\ or \textquotedblleft phase\textquotedblright\
variability (e.g.~variability in the location of the mass peaks in Figure %
\ref{fig:curves}(a)) and \textquotedblleft vertical\textquotedblright\ or
\textquotedblleft amplitude\textquotedblright\ variability (e.g.~variability
in the mass magnitude at the peak in Figure \ref{fig:curves}(a)). It is
important to point out that for a given data set there is often some
ambiguity about what constitutes amplitude variability and what constitutes
phase variability (this will be discussed in more depth in Section \ref%
{sec:background}). The problem of decomposing functional variability into
amplitude and phase variability has been addressed by many authors (Kneip
and Engel, 1995; Ramsay and Li, 1998; Wang and Gasser, 1999; Kneip et al.,
2000; Gervini and Gasser, 2004, 2005; Kneip and Ramsay, 2008; Tang and M\"{u}%
ller, 2008; Telesca and Inoue, 2008; Bigot and Gadat, 2010; Claeskens et
al., 2010). All of these papers, however, have focused on independent and
identically distributed samples of curves, but for the type of applications
we have in mind the curves are not independent. For example, the growth
curves in Figure \ref{fig:curves}(a) are correlated for individuals with the
same father. This type of design is common in evolutionary biology and
quantitative genetics for the following reason. The variability observed in
physical traits has two sources: genetic and environmental. Because
environmental factors generally are not passed from one generation to the
next, the evolution of phenotypes is driven largely by genetic variability
(but see Skinner et al., 2010 and Manikkam et al., 2012 for a discussion of
epigenetic effects). Examining samples of genetically related individuals,
like siblings or half-siblings, makes the genetic and environmental sources
of variability mathematically identifiable and therefore estimable, allowing
biologists to predict the evolution of traits in response to selection
(Gomulkiewicz and Beder, 1996; Kingsolver at al., 2002).

Therefore, it is important to possess statistical tools for the study of
amplitude and phase variability of non-independent functional data. Some
existing functional-data methods handle non-independent or non-identically
distributed curves, such as mixed-effects ANOVA models (Guo, 2002; Morris
and Carroll, 2006; Di et al., 2009; Chen and Wang, 2011), but they do not
specifically address phase variability. To date, the problem of
amplitude/phase variability of functional traits has been addressed mostly
in an ad-hoc way, by first aligning the curves with respect to some trait,
and then studying amplitude variability of the aligned curves. (This process
of aligning curves is variously known as \textquotedblleft time
warping\textquotedblright\ or \textquotedblleft curve
registration\textquotedblright\ in the Functional Data literature.) But
evolutionary biologists frequently must make decisions about how to align or
register individual curves from a population of individuals. For example,
when studying growth curves in a population of animals that undergo
metamorphosis from one life history state to another (usually from a
non-reproductive larval form to a reproductive adult form), it is not
necessarily clear how to align the individual curves. The default choice for
most biologists is to align the curves at the date of birth or hatching, but
an equally valid choice might be the date of metamorphosis or the peak body
mass prior to metamorphosis. For example, Ragland and Carter (2004) chose to
align the growth curves of larval salamanders by date of metamorphosis and
then reset the growth period to a fractional scale. Although this approach
was effective, it was unsophisticated and ad hoc; more rigorous methods
would be beneficial.

In this paper we propose a functional ANOVA approach that explicitly models
time variability. For simplicity, we consider only the one-way random factor
model, but the ideas can be easily extended to more complex ANOVA models. We
follow a likelihood-based approach that uses the raw data directly, without
pre-smoothing. Therefore the method can be applied to irregularly sampled
trajectories, with possibly different starting points and endpoints. The
fact that pre-smoothing is not necessary makes the method applicable to
longitudinal data, where a smooth latent process is assumed but the observed
data themselves are not smooth (Rice, 2004; M\"{u}ller, 2008). The paper is
organized as follows: a brief background on random processes is given in
Section \ref{sec:background}; the warped ANOVA model is presented in Section %
\ref{sec:Model}; the asymptotic distribution of the main parameter
estimators is derived in Section \ref{sec:Inference}; the small sample
behavior of the estimators is studied by simulation in Section \ref%
{sec:Simulations}; finally, the beetle growth data is analyzed in detail in
Section \ref{sec:Example}.

\section{\label{sec:background}Brief background on random processes}

Before we present the warped ANOVA model, it is useful to review some basic
properties of stochastic processes. Let $x:I\rightarrow \mathbb{R}$ be a
random function defined on a finite interval $I\subset \mathbb{R}$. Suppose $%
x(t)$ is square-integrable with probability one, and has finite variance.
Let $\mu (t)=\mathrm{E}\{x(t)\}$ and $\rho (s,t)=\mathrm{cov}\{x(s),x(t)\}$.
Then $x(t)$ admits the decomposition 
\begin{equation}
x(t)=\mu (t)+\sum_{k=1}^{\infty }Z_{k}\phi _{k}(t),  \label{eq:KL_decomp}
\end{equation}%
which is known as Karhunen--Lo\`{e}ve decomposition (Ash and Gardner, 1975),
where the $Z_{k}$s are uncorrelated random variables with $\mathrm{E}%
(Z_{k})=0$ and $\mathrm{var}(Z_{k})=\lambda _{k}$ (without loss of
generality we can assume $\lambda _{1}\geq \lambda _{2}\geq \cdots \geq 0$).
The $\phi _{k}$s form an orthonormal system in $\mathcal{L}^{2}(I)$ and are
eigenfunctions of the covariance function $\rho $ with eigenvalues $\lambda
_{k}$; that is, $\int \rho (s,t)\phi _{k}(s)ds=\lambda _{k}\phi _{k}(t)$,
which implies 
\begin{equation}
\rho (s,t)=\sum_{k=1}^{\infty }\lambda _{k}\phi _{k}(s)\phi _{k}(t).
\label{eq:spectral_decomp}
\end{equation}%
If the covariance function $\rho $ is continuous then (\ref{eq:KL_decomp})
and (\ref{eq:spectral_decomp}) converge pointwise; otherwise the convergence
is only in the sense of the $\mathcal{L}^{2}(I)$ norm (Gohberg \emph{et al.}%
, 2003). In either case, $\sum_{k=1}^{\infty }\lambda _{k}<\infty $, so the
sequence of eigenvalues converges to zero. The Karhunen--Lo\`{e}ve
decomposition is the functional equivalent of the multivariate
principal-component decomposition.

Although from a mathematical point of view decomposition (\ref{eq:KL_decomp}%
) always holds, from a statistical point of view it is not always the most
parsimonious model. It is often the case that the sample curves present a
few distinct peaks and valleys that systematically repeat themselves across
curves, albeit at somewhat different locations. It may take a lot of terms
in (\ref{eq:KL_decomp}) to explain this kind of variability, but Kneip and
Ramsay (2008, Proposition 1) show that if the process $x(t)$ has at most $K$
peaks and valleys and its derivative $x^{\prime }(t)$ has at most $K$ zeros,
then $x(t)$ admits the decomposition 
\begin{equation}
x(t)=\sum_{j=1}^{p}C_{j}\xi _{j}\{v(t)\}  \label{eq:Kneip_exp}
\end{equation}%
for some $p\leq K+2$, where the $\xi _{j}$s are non-random basis functions,
the $C_{j}$s are random coefficients, and $v:I\rightarrow I$ is a monotone
increasing stochastic process such that $\mathrm{E}\{v(t)\}=t$ (or
alternatively $\mathrm{E}\{w(s)\}=s$, where $w(t)$ is the inverse function
of $v(t)$.) We can re-express (\ref{eq:Kneip_exp}) as 
\begin{equation}
x\{w(s)\}=\mu ^{\ast }(s)+\sum_{k=1}^{p}Z_{k}^{\ast }\phi _{k}^{\ast }(s),
\label{eq:warped_KL_decomp}
\end{equation}%
which is just the Karhunen--Lo\`{e}ve decomposition of the warped process $%
z=x\circ w$. The process $w$ is called the warping process, and it explains
the \textquotedblleft horizontal\textquotedblright\ variability in the
location of the peaks and valleys of $x$.

It is important to point out that the Karhunen--Lo\`{e}ve decomposition (\ref%
{eq:KL_decomp}) is essentially unique (up to the usual indeterminacy of
eigenfunctions for multiple eigenvalues), and so is (\ref%
{eq:warped_KL_decomp}) for a given warping process $w$; but the warping
process $w$ itself is not unique. For a given $x(t)$, different warping
processes $w(t)$ can be chosen that will give rise to different
decompositions (\ref{eq:warped_KL_decomp}). In general, it is not possible
to uniquely define what constitutes amplitude\ variability and what
constitutes phase\ variability for a given process $x(t)$. The approach
usually followed in the literature is to specify a warping family $\mathcal{W%
}$ where $w(t)$ is constrained to live, and then define as phase variability
whatever is accounted for by the family $\mathcal{W}$ and as amplitude
variability whatever is accounted for by the residual decomposition (\ref%
{eq:warped_KL_decomp}). This may sound vague, but in fact it is possible to
give simple conditions for model (\ref{eq:warped_KL_decomp}) to be
identifiable given a family $\mathcal{W}$; see the discussion in Web
Appendix E. Some authors choose very rigid warping families $\mathcal{W}$,
like linear warping functions (Sangalli \emph{et al.}, 2010), while others
use extremely flexible nonparametric families (Telesca and Inoue, 2008;
Ramsay and Li, 1998). We will follow an intermediate approach, using the
semiparametric family of interpolating monotone Hermite splines (Fritsch and
Carlson, 1980), although the proposed method can be implemented with any
other warping family.

Monotone interpolating Hermite splines are defined as follows (more details
are given in Web Appendix C). For a subject $i$, let $\mathbf{\tau }_{i}\in 
\mathbb{R}^{r}$ be a vector of \textquotedblleft
landmarks\textquotedblright\ in $I=[a,b]$, with $a<\tau _{i1}<\cdots <\tau
_{ir}<b$; the $\tau $s can be, for example, the locations of the peaks and
valleys of the observed curve. Let $\mathbf{\tau }_{0}\in \mathbb{R}^{r}$ be
a knot vector, usually taken as the mean of the $\mathbf{\tau }_{i}$s. For
given values $s_{i0},\ldots ,s_{i,r+1}$, there exists a unique piecewise
cubic function $w_{i}(t)$ such that $w_{i}(a)=a$, $w_{i}(b)=b$, $w_{i}(\tau
_{0j})=\tau _{ij}$ for all $j$, $w_{i}^{\prime }(a)=s_{i0}$, $w_{i}^{\prime
}(b)=s_{i,r+1}$, and $w_{i}^{\prime }(\tau _{0j})=s_{ij}$ for all $j$. This
function $w_{i}(t)$ aligns the individual features $\mathbf{\tau }_{i}$ with
the average features $\mathbf{\tau }_{0}$ in a smooth way, so it is useful
for \textquotedblleft landmark registration\textquotedblright\ (Bookstein,
1997). For $w_{i}(t)$ to be strictly monotone increasing the derivatives $%
s_{ij}$s must satisfy certain conditions, given in Fritsch and Carlson
(1980). But for curve-alignment purposes only the $\mathbf{\tau }_{i}$s are
specified; in that case Fritsch and Carlson (1980) provide an algorithm that
produces a vector of derivatives $\mathbf{s}_{i}$ that satisfy the
sufficient conditions for $w_{i}(t)$ to be monotone increasing. Since the
algorithm is deterministic, $\mathbf{s}_{i}$ is a function of $\mathbf{\tau }%
_{i}$ and $\mathbf{\tau }_{0}$, therefore $w_{i}(t)$ is entirely
parameterized by $\mathbf{\tau }_{i}$ and $\mathbf{\tau }_{0}$. In this
paper, instead of specifying $\mathbf{\tau }_{i}$ for each curve and taking $%
\mathbf{\tau }_{0}=\mathbf{\bar{\tau}}$, we will specify $\mathbf{\tau }_{0}$
and treat the $\mathbf{\tau }_{i}$s as unobserved random effects. Our family
of warping functions $\mathcal{W}_{\mathbf{\tau }_{0}}$, then, is an $r$%
-dimensional space ($r$ will usually be small). In general, it is not
problematic to specify a reasonable $\mathbf{\tau }_{0}$ for a given data
set; for example, for the curves in Figure \ref{fig:curves}(a) a single knot
at $\tau _{0}=15$ will provide reasonable warping flexibility, and the rest
of the variation will be considered amplitude variability. For other warping
families, such as monotone B-splines (Telesca and Inoue, 2008) or smooth
monotone transformations (Ramsay and Li, 1998), the number and placement of
the knots are harder to specify because they are not directly associated
with curve features.

\section{\label{sec:Model}The warped ANOVA model}

Let us go back now to the original problem of a one-factor design, where the
sample of $n$ individuals can be separated into $I$ groups, with group $i$
containing $J_{i}$ individuals. For subject $j$ in group $i$ we observe
certain variable (e.g.~mass) at time points $t_{ij1},\ldots ,t_{ij\nu _{ij}}$%
, obtaining observations $y_{ij1},\ldots ,y_{ij\nu _{ij}}$. The number of
observations $\nu _{ij}$ as well as the time points may change from
individual to individual. We assume 
\begin{equation}
y_{ijk}=x_{ij}(t_{ijk})+\varepsilon _{ijk},  \label{eq:raw_data_model}
\end{equation}%
where $\{x_{ij}(t)\}$ are underlying smooth curves, no directly observable,
and $\{\varepsilon _{ijk}\}$ are i.i.d.$~N(0,\sigma ^{2})$ random errors
independent of the underlying $x_{ij}(t)$s. Observational model (\ref%
{eq:raw_data_model}), which treats the smooth curves $\{x_{ij}(t)\}$ as
latent variables, is the usual way to bridge functional data analysis and
longitudinal data analysis (M\"{u}ller, 2008). As discussed in Section \ref%
{sec:background}, we can write $x_{ij}(t)=z_{ij}\{w_{ij}^{-1}(t)\}$ for a
warped process $z_{ij}(t)$ and a warping function $w_{ij}(t)$. These will
inherit the dependence structure of the $x_{ij}$s, so we can assume 
\begin{equation}
z_{ij}(t)=\mu (t)+\alpha _{i}(t)+\beta _{ij}(t),\ \ j=1,\ldots ,J_{i},\ \
i=1,\ldots ,I,  \label{eq:z_model}
\end{equation}%
with $\{\alpha _{i}(t)\}$ and $\{\beta _{ij}(t)\}$ zero-mean random factors
independent of each other and among themselves. For the main factor $\alpha
(t)$ and the residual term $\beta (t)$ we assume expansions analogous to (%
\ref{eq:warped_KL_decomp}): 
\begin{equation}
\alpha (t)=\sum_{k=1}^{p}U_{k}\phi _{k}(t),  \label{eq:KL_alpha}
\end{equation}%
\begin{equation}
\beta (t)=\sum_{k=1}^{q}V_{k}\psi _{k}(t),  \label{eq:KL_beta}
\end{equation}%
where $\{\phi _{k}(t)\}$ and $\{\psi _{k}(t)\}$ are orthonormal functions in 
$\mathcal{L}^{2}(I)$, the $U_{k}$s are uncorrelated with $\mathrm{E}%
(U_{k})=0 $ and $\mathrm{var}(U_{k})=\gamma _{k}$, and the $V_{k}$s are
uncorrelated with $\mathrm{E}(V_{k})=0$ and $\mathrm{var}(V_{k})=\lambda
_{k} $. Without loss of generality we assume $\gamma _{1}\geq \cdots \geq
\gamma _{p}>0$ and $\lambda _{1}\geq \cdots \geq \lambda _{q}>0$.

From (\ref{eq:z_model}), (\ref{eq:KL_alpha}) and (\ref{eq:KL_beta}) it
follows that the total variance of $z_{ij}(t)$, defined as $\mathrm{E}(\Vert
z_{ij}-\mu \Vert ^{2})$ with $\left\Vert \cdot \right\Vert $ the usual $%
\mathcal{L}^{2}$-norm, can be decomposed as $\mathrm{E}(\Vert \alpha \Vert
^{2})+\mathrm{E}(\Vert \beta \Vert ^{2})$, where $\mathrm{E}(\Vert \alpha
\Vert ^{2})=\sum_{k=1}^{p}\gamma _{k}$ is the main-factor variance and $%
\mathrm{E}(\Vert \beta \Vert ^{2})=\sum_{k=1}^{q}\lambda _{k}$ is the
residual-factor variance. The ratio 
\begin{equation}
h_{z}=\frac{\sum_{k=1}^{p}\gamma _{k}}{\sum_{k=1}^{p}\gamma
_{k}+\sum_{k=1}^{q}\lambda _{k}}  \label{eq:h_z}
\end{equation}%
is then the proportion of amplitude variability explained by the main
factor. In Section \ref{sec:Inference} we will derive asymptotic confidence
intervals for $h_{z}$.

The mean function $\mu (t)$ and the Karhunen--Lo\`{e}ve components $\{\phi
_{k}(t)\}$ and $\{\psi _{k}(t)\}$ are functional parameters that must be
estimated from the data, using for instance semiparametric spline models.
Let $\mathbf{b}(t)=(b_{1}(t),\ldots ,b_{s}(t))^{T}$ be a spline basis in $%
\mathcal{L}^{2}(I)$ (for simplicity we will use the same spline basis for
all functional parameters, but this is not strictly necessary); then we
assume $\mu (t)=\mathbf{b}(t)^{T}\mathbf{m}$, $\phi _{k}(t)=\mathbf{b}(t)^{T}%
\mathbf{c}_{k}$, and $\psi _{k}(t)=\mathbf{b}(t)^{T}\mathbf{d}_{k}$, for
parameters $\mathbf{m}$, $\mathbf{c}_{k}$ and $\mathbf{d}_{k}$ in $\mathbb{R}%
^{s}$. Let $\mathbf{C=[c}_{1},\ldots ,\mathbf{c}_{p}]\in \mathbb{R}^{s\times
p}$, $\mathbf{D=[d}_{1},\ldots ,\mathbf{d}_{q}]\in \mathbb{R}^{s\times q}$
and $\mathbf{J}=\int_{a}^{b}\mathbf{b}(t)\mathbf{b}(t)^{T}dt$ $\in \mathbb{R}%
^{s\times s}$. The orthogonality conditions on the $\phi _{k}$s and the $%
\psi _{k}$s translate into the conditions $\mathbf{C}^{T}\mathbf{JC}=\mathbf{%
I}_{p}$ and $\mathbf{D}^{T}\mathbf{JD}=\mathbf{I}_{q}$ for $\mathbf{C}$ and $%
\mathbf{D}$. Regarding the $U_{k}$s and $V_{k}$s in (\ref{eq:KL_alpha}) and (%
\ref{eq:KL_beta}), we assume that $\mathbf{U}=(U_{1},\ldots ,U_{p})^{T}$
follows a multivariate $N(\mathbf{0},\mathbf{\Gamma })$ distribution with $%
\mathbf{\Gamma }=\mathrm{diag}(\gamma _{1},\ldots ,\gamma _{p})$ and that $%
\mathbf{V}=(V_{1},\ldots ,V_{q})^{T}$ follows a multivariate $N(\mathbf{0},%
\mathbf{\Lambda })$ distribution with $\mathbf{\Lambda }=\mathrm{diag}%
(\lambda _{1},\ldots ,\lambda _{q})$. To summarize, the warped process (\ref%
{eq:z_model}) is parameterized by $\mathbf{m}$, $\mathbf{C}$, $\mathbf{D}$, $%
\mathbf{\Gamma }$ and $\mathbf{\Lambda }$.

For the warping functions $w_{ij}(t)$ we cannot simply assume an additive
model like (\ref{eq:z_model}) with Gaussian factors, because there are no
monotone Gaussian processes. Therefore, a more indirect approach is needed.
We will assume the $w_{ij}$s belong to the family of interpolating Hermite
cubic splines $\mathcal{W}_{\mathbf{\tau }_{0}}$ for a specified knot vector 
$\mathbf{\tau }_{0}$. We have seen in Section \ref{sec:background} that a $%
w_{ij}\in \mathcal{W}_{\mathbf{\tau }_{0}}$ is parameterized by a vector $%
\mathbf{\tau }_{ij}$ that can be treated as a random effect. However, due to
the restriction $a<\tau _{ij1}<\cdots <\tau _{ijr}<b$ we cannot assume $%
\mathbf{\tau }_{ij}$ is Normal. So we follow the approach of Brumback and
Lindstrom (2004) and use the Jupp (1978) transform $\mathbf{\theta }_{ij}=%
\mathcal{J}(\mathbf{\tau }_{ij})$, defined as $\theta _{ijk}=\log \{(\tau
_{ij,k+1}-\tau _{ijk})/(\tau _{ijk}-\tau _{ij,k-1})\}$ for $k=1,\ldots ,r$,
which is an invertible transformation that maps vectors $\mathbf{\tau }_{ij}$
with increasing coordinates into unconstrained vectors $\mathbf{\theta }%
_{ij} $. For the unconstrained vector $\mathbf{\theta }_{ij}$ we can assume
a multivariate Normal distribution and an additive ANOVA model: 
\begin{equation}
\mathbf{\theta }_{ij}=\mathbf{\theta }_{0}+\mathbf{\eta }_{i}+\mathbf{\xi }%
_{ij},  \label{eq:model_th}
\end{equation}%
with $\mathbf{\eta }_{i}\sim N(\mathbf{0},\mathbf{\Sigma })$ and $\mathbf{%
\xi }_{ij}\sim N(\mathbf{0},\mathbf{\Omega })$ independent of each other and
among themselves. We will also assume the $\mathbf{\theta }_{ij}$s are
independent of the amplitude factors $\alpha _{i}(t)$ and $\beta _{ij}(t)$,
although a model with correlations between amplitude and warping factors can
be set up (see below). We take $\mathbf{\theta }_{0}=\mathcal{J}(\mathbf{%
\tau }_{0})$; the covariance matrices $\mathbf{\Sigma }$ and $\mathbf{\Omega 
}$ will be estimated from the data. In analogy with (\ref{eq:h_z}) we define 
\begin{equation}
h_{w}=\frac{\mathrm{tr}(\mathbf{\Sigma })}{\mathrm{tr}(\mathbf{\Sigma }+%
\mathbf{\Omega })},  \label{eq:h_w}
\end{equation}%
which is the proportion of the warping variability explained by the main
factor.

Putting together the models for $z_{ij}(t)$, $w_{ij}(t)$ and the
observational model (\ref{eq:raw_data_model}), we can derive the likelihood
function for the observed data vectors $\mathbf{y}_{ij}=(y_{ij1},\ldots
,y_{ij\nu _{ij}})$. Given a realization of the random effect $\mathbf{\theta 
}_{ij}$, which is determined by realizations of $\mathbf{\eta }_{i}$ and $%
\mathbf{\xi }_{ij}$, the corresponding warped time grids are $t_{ijk}^{\ast
}(\mathbf{\theta }_{ij})=w_{ij}^{-1}(t_{ijk})$, $k=1,\ldots ,\nu _{ij}$, and
the corresponding warped B-spline matrices $\mathbf{B}_{ij}^{\ast }(\mathbf{%
\theta }_{ij})\in \mathbb{R}^{\nu _{ij}\times s}$ are given by $[\mathbf{B}%
_{ij}^{\ast }(\mathbf{\theta }_{ij})]_{kl}=b_{l}\{t_{ijk}^{\ast }(\mathbf{%
\theta }_{ij})\}$. Then 
\begin{equation*}
\mathbf{y}_{ij}|(\mathbf{u}_{i},\mathbf{v}_{ij}\mathbf{,\eta }_{i},\mathbf{%
\xi }_{ij})\sim N\left\{ \mathbf{B}_{ij}^{\ast }(\mathbf{\theta }_{ij})%
\mathbf{m+B}_{ij}^{\ast }(\mathbf{\theta }_{ij})\mathbf{Cu}_{i}\mathbf{+B}%
_{ij}^{\ast }(\mathbf{\theta }_{ij})\mathbf{Dv}_{ij},\sigma ^{2}\mathbf{I}%
_{\nu _{ij}}\right\} ,
\end{equation*}%
and the $\mathbf{y}_{ij}$s are conditionally independent given $(\mathbf{u}%
_{i},\mathbf{v}_{ij}\mathbf{,\eta }_{i},\mathbf{\xi }_{ij})$. If $\mathbf{y}%
_{i\cdot }=(\mathbf{y}_{i1},\ldots ,\mathbf{y}_{iJ_{i}})$, we have 
\begin{equation}
f(\mathbf{y}_{i\cdot })=\iint g(\mathbf{u}_{i},\mathbf{\eta }_{i})f(\mathbf{u%
}_{i})f(\mathbf{\eta }_{i})\mathrm{d}\mathbf{u}_{i}\mathrm{d}\mathbf{\eta }%
_{i}  \label{eq:f_y}
\end{equation}%
with $g(\mathbf{u}_{i},\mathbf{\eta }_{i})=\prod_{j=1}^{J_{i}}\iint f(%
\mathbf{y}_{ij}|\mathbf{u}_{i},\mathbf{v}_{ij}\mathbf{,\eta }_{i},\mathbf{%
\xi }_{ij})f(\mathbf{v}_{ij})f(\mathbf{\xi }_{ij})\mathrm{d}\mathbf{v}_{ij}%
\mathrm{d}\mathbf{\xi }_{ij}$, and the log-likelihood function is $\ell
=\sum_{i=1}^{I}\log f(\mathbf{y}_{i\cdot })$. The maximum likelihood
estimators are $(\mathbf{\hat{m}},\mathbf{\hat{C}},\mathbf{\hat{D}},\mathbf{%
\hat{\Lambda}},\mathbf{\hat{\Gamma}},\allowbreak \mathbf{\hat{\Sigma}},%
\mathbf{\hat{\Omega}},\hat{\sigma}^{2})=\arg \max \ell $. We compute them
via the EM algorithm. The implementation of the EM algorithm presents
certain complications arising from the orthogonality restrictions on $%
\mathbf{C}$ and $\mathbf{D}$, which are discussed in detail in Web Appendix
A.

In the rest of the paper we will use the estimators as presented above, but
to conclude the section we discuss a few possible generalizations. First, it
is possible to use other families of warping functions, such as B-splines
with monotone increasing coefficients (Brumback and Lindstrom, 2004; Telesca
and Inoue, 2008) or smooth monotone transformations (Ramsay and Li, 1998).
The problem is that the spline coefficients for these families cannot be
directly related to features of the sample curves in the way interpolating
Hermite spline coefficients can; therefore, one may have to use a relatively
large number of knots placed at somewhat arbitrary locations (equally
spaced, for example). This may result in a warping family that is too
flexible and lead to overwarping (i.e., produce warping functions with flat
parts that are close to singular). To prevent this, the warping variability
must somehow be penalized. This can be done by adding a penalty term to the
log-likelihood function and minimize $\ell _{\lambda }=\sum_{i=1}^{I}\log f(%
\mathbf{y}_{i\cdot })-\lambda \mathrm{tr}(\mathbf{\Sigma }+\mathbf{\Omega })$%
, where $\lambda \geq 0$ is a penalty parameter chosen by the user.

Second, it is possible to incorporate correlations between the warping
process $w_{ij}(t)$ and the amplitude process $z_{ij}(t)$. This can be done
by assuming that $(\mathbf{u}_{i},\mathbf{\eta }_{i})$ and $(\mathbf{v}_{ij},%
\mathbf{\xi }_{ij})$ have joint Normal distributions, for instance. The only
change in (\ref{eq:f_y}) would be that $f(\mathbf{u}_{i})f(\mathbf{\eta }%
_{i})$ is replaced by the joint density $f(\mathbf{u}_{i},\mathbf{\eta }%
_{i}) $ and $f(\mathbf{v}_{ij})f(\mathbf{\xi }_{ij})$ by $f(\mathbf{v}_{ij},%
\mathbf{\xi }_{ij})$. From the computational point of view this does not
have a big impact, because the EM algorithm can be easily modified to
accommodate this (a Matlab implementation is available as supplementary
material). But from a statistical point of view the results may be harder to
interpret, and the extra $rp+rq$ covariance parameters that need to be
estimated may affect the precision of the rest of the estimators if the
sample size is not very large.

Finally, we note that the assumption of normality of the random effects is
mostly a working assumption to derive estimators. It is usually the case
that properties like consistency and asymptotic normality of maximum
likelihood estimators hold for broader families of distributions than the
one they were derived for. But such a thorough asymptotic analysis is beyond
the scope of this paper. Instead, in Section \ref{sec:Simulations} we will
study by simulation the robustness of the estimators to at least some mild
departures from normality. On the other hand, if robustness to outliers is
desired, this may be attained by substituting the Normal distributions by
multivariate $t$ distributions, as in Gervini (2009); the Normal EM
algorithm is easy to adapt for multivariate $t$ distributions.

\section{\label{sec:Inference}Asymptotics and inference}

It is usually of interest in applications to determine if the main-factor
variance is significantly different from zero or not. To this end, we derive
in this section the asymptotic distributions of the maximum likelihood
estimators and the variance ratios (\ref{eq:h_z}) and (\ref{eq:h_w}), which
can then be used to construct asymptotic confidence intervals and tests for $%
h_{z}$ and $h_{w}$. For simplicity, we assume that \emph{(i)} the true
functional parameters $\mu (t)$, $\{\phi _{k}(t)\}$ and $\{\psi _{k}(t)\}$
belong to the spline space used for estimation, which is fixed, and \emph{%
(ii)} the $\mathbf{y}_{i\cdot }$s are identically distributed, so $J_{i}=J$
for all $i$ and the time grid $(t_{1},\ldots ,t_{\nu })$ is the same for all
individuals. The asymptotic distribution of the estimators will be derived
for $I\rightarrow \infty $ and $J$ fixed, or in practical terms, for
\textquotedblleft large\ $I$ and small\ $J$\textquotedblright ; this is the
usual situation in random-effect one-way ANOVA models.

Under these conditions the standard maximum likelihood asymptotic theory
applies: if $\mathbf{\omega }=(\gamma _{1},\ldots ,\gamma _{p},\lambda
_{1},\ldots ,\lambda _{q})$, then $\sqrt{I}(\mathbf{\hat{\omega}}-\mathbf{%
\omega })\overset{D}{\longrightarrow }N(\mathbf{0},\mathbf{F}^{-1})$, where $%
\mathbf{F}=\mathrm{E}[\{\frac{\partial }{\partial \mathbf{\omega }}\log f(%
\mathbf{y}_{i\cdot })\}\allowbreak \{\frac{\partial }{\partial \mathbf{%
\omega }}\log f(\mathbf{y}_{i\cdot })\}^{T}]$ is the Fisher Information
Matrix for the parameter $\mathbf{\omega }$. Straightforward differentiation
of (\ref{eq:f_y}), which is carried out in detail in Web Appendix B, gives 
\begin{equation*}
\frac{\partial }{\partial \gamma _{k}}\log f(\mathbf{y}_{i\cdot })=-\frac{1}{%
2\gamma _{k}}+\frac{\mathrm{E}(u_{ik}^{2}|\mathbf{y}_{i\cdot })}{2\gamma
_{k}^{2}},\ \ k=1,\ldots ,p,
\end{equation*}%
and 
\begin{equation*}
\frac{\partial }{\partial \lambda _{k}}\log f(\mathbf{y}_{i\cdot })=-\frac{J%
}{2\lambda _{k}}+\frac{1}{2\lambda _{k}^{2}}\sum_{j=1}^{J}\mathrm{E}%
(v_{ijk}^{2}|\mathbf{y}_{i\cdot }),\ \ k=1,\ldots ,q.
\end{equation*}%
Let $\widehat{u_{ik}^{2}}=\mathrm{E}(u_{ik}^{2}|\mathbf{y}_{i\cdot })$ and $%
\widehat{v_{ijk}^{2}}=\mathrm{E}(v_{ijk}^{2}|\mathbf{y}_{i\cdot })$. Since $%
\mathrm{E}(\widehat{u_{ik}^{2}})=\mathrm{E}(u_{ik}^{2})=\gamma _{k}$ and $%
\mathrm{E}(\widehat{v_{ijk}^{2}})=\mathrm{E}(v_{ijk}^{2})=\lambda _{k}$, we
obtain the following expressions:

\begin{equation*}
F_{kl}=-\frac{1}{4\gamma _{k}\gamma _{l}}+\frac{\mathrm{E}(\widehat{%
u_{ik}^{2}}\widehat{u_{il}^{2}})}{4\gamma _{k}^{2}\gamma _{l}^{2}},\ \ \text{%
for }k=1,\ldots ,p\text{ and }l=1,\ldots ,p,
\end{equation*}%
\begin{equation*}
F_{k,p+l}=-\frac{J}{4\gamma _{k}\lambda _{l}}+\frac{\mathrm{E}(\widehat{%
u_{ik}^{2}}\sum_{j=1}^{J}\widehat{v_{ijl}^{2}})}{4\gamma _{k}^{2}\lambda
_{l}^{2}},\ \ \text{for }k=1,\ldots ,p\text{ and }l=1,\ldots ,q,
\end{equation*}%
and 
\begin{equation*}
F_{p+k,p+l}=-\frac{J^{2}}{4\lambda _{k}\lambda _{l}}+\frac{\mathrm{E}%
(\sum_{j=1}^{J}\widehat{v_{ijk}^{2}}\sum_{j=1}^{J}\widehat{v_{ijl}^{2}})}{%
4\lambda _{k}^{2}\lambda _{l}^{2}},\ \ \text{for }k=1,\ldots q\text{ and }%
l=1,\ldots ,q.
\end{equation*}%
The estimator $\mathbf{\hat{F}}$ is obtained replacing expectations by
averages over $i=1,\ldots ,I$.

The asymptotic distribution of (\ref{eq:h_z}) is derived via the Delta
Method: since $h_{z}$ is a differentiable function of $\mathbf{\omega }$, $%
\sqrt{I}(\hat{h}_{z}-h_{z})\overset{D}{\longrightarrow }N\left( 0,\left\{
\partial h_{z}/\partial \mathbf{\omega }\right\} ^{T}\mathbf{F}^{-1}\left\{
\partial h_{z}/\partial \mathbf{\omega }\right\} \right) $ with 
\begin{equation*}
\frac{\partial h_{z}}{\partial \gamma _{k}}=\frac{\sum_{k=1}^{q}\lambda _{k}%
}{\left( \sum_{k=1}^{p}\gamma _{k}+\sum_{k=1}^{q}\lambda _{k}\right) ^{2}},\
\ k=1,\ldots ,p,
\end{equation*}%
and 
\begin{equation*}
\frac{\partial h_{z}}{\partial \lambda _{k}}=-\frac{\sum_{k=1}^{p}\gamma _{k}%
}{\left( \sum_{k=1}^{p}\gamma _{k}+\sum_{k=1}^{q}\lambda _{k}\right) ^{2}},\
\ k=1,\ldots ,q.
\end{equation*}%
The asymptotic variance of $\hat{h}_{z}$ is then given by 
\begin{eqnarray*}
\mathrm{avar}(\hat{h}_{z}) &=&\frac{\left( \sum_{k=1}^{p}\lambda _{k}\right)
^{2}}{\left( \sum_{k=1}^{p}\gamma _{k}+\sum_{k=1}^{q}\lambda _{k}\right) ^{4}%
}\cdot \sum_{k=1}^{p}\sum_{l=1}^{p}\left( \mathbf{F}^{-1}\right) _{kl} \\
&&-\frac{2\left( \sum_{k=1}^{p}\gamma _{k}\right) \left(
\sum_{k=1}^{q}\lambda _{k}\right) }{\left( \sum_{k=1}^{p}\gamma
_{k}+\sum_{k=1}^{q}\lambda _{k}\right) ^{4}}\cdot
\sum_{k=1}^{p}\sum_{l=1}^{q}\left( \mathbf{F}^{-1}\right) _{k,p+l} \\
&&+\frac{\left( \sum_{k=1}^{q}\gamma _{k}\right) ^{2}}{\left(
\sum_{k=1}^{p}\gamma _{k}+\sum_{k=1}^{q}\lambda _{k}\right) ^{4}}\cdot
\sum_{k=1}^{q}\sum_{l=1}^{q}\left( \mathbf{F}^{-1}\right) _{p+k,p+l}.
\end{eqnarray*}

The asymptotic distribution of (\ref{eq:h_w}) is derived in a similar way.
If $\mathbf{\zeta }=(\mathrm{diag}(\mathbf{\Sigma }),\mathrm{diag}(\mathbf{%
\Omega }))$, then $\sqrt{I}(\mathbf{\hat{\zeta}}-\mathbf{\zeta })\overset{D}{%
\longrightarrow }N(\mathbf{0},\mathbf{G}^{-1})$ with $\mathbf{G}=\mathrm{E}%
[\{\frac{\partial }{\partial \mathbf{\zeta }}\log f(\mathbf{y}_{i\cdot })\}\{%
\frac{\partial }{\partial \mathbf{\zeta }}\log f(\mathbf{y}_{i\cdot
})\}^{T}] $, and $\sqrt{I}(\hat{h}_{w}-h_{w})\overset{D}{\longrightarrow }%
N(0,\mathrm{avar}(\hat{h}_{w}))$ with $\mathrm{avar}(\hat{h}_{w})=\left(
\partial h_{w}/\partial \mathbf{\zeta }\right) ^{T}\mathbf{G}^{-1}\left(
\partial h_{w}/\partial \mathbf{\zeta }\right) $, where 
\begin{equation*}
\frac{\partial h_{w}}{\partial \Sigma _{kk}}=\frac{\mathrm{tr}(\mathbf{%
\Omega })}{\left\{ \mathrm{tr}(\mathbf{\Sigma })+\mathrm{tr}(\mathbf{\Omega }%
)\right\} ^{2}},\ \ \text{for }k=1,\ldots ,r,
\end{equation*}%
and 
\begin{equation*}
\frac{\partial h_{w}}{\partial \Omega _{kk}}=-\frac{\mathrm{tr}(\mathbf{%
\Sigma })}{\left\{ \mathrm{tr}(\mathbf{\Sigma })+\mathrm{tr}(\mathbf{\Omega }%
)\right\} ^{2}},\ \ \text{for }k=1,\ldots ,r.
\end{equation*}%
Then 
\begin{eqnarray*}
\mathrm{avar}(\hat{h}_{w}) &=&\frac{\left\{ \mathrm{tr}(\mathbf{\Omega }%
)\right\} ^{2}}{\left\{ \mathrm{tr}(\mathbf{\Sigma })+\mathrm{tr}(\mathbf{%
\Omega })\right\} ^{4}}\cdot \sum_{k=1}^{r}\sum_{l=1}^{r}\left( \mathbf{G}%
^{-1}\right) _{kl} \\
&&-\frac{2\mathrm{tr}(\mathbf{\Sigma })\mathrm{tr}(\mathbf{\Omega })}{%
\left\{ \mathrm{tr}(\mathbf{\Sigma })+\mathrm{tr}(\mathbf{\Omega })\right\}
^{4}}\cdot \sum_{k=1}^{r}\sum_{l=1}^{r}\left( \mathbf{G}^{-1}\right) _{k,r+l}
\\
&&+\frac{\left\{ \mathrm{tr}(\mathbf{\Sigma })\right\} ^{2}}{\left\{ \mathrm{%
tr}(\mathbf{\Sigma })+\mathrm{tr}(\mathbf{\Omega })\right\} ^{4}}\cdot
\sum_{k=1}^{r}\sum_{l=1}^{r}\left( \mathbf{G}^{-1}\right) _{r+k,r+l}.
\end{eqnarray*}%
As shown in Web Appendix B, differentiation of (\ref{eq:f_y}) gives 
\begin{equation*}
\frac{\partial }{\partial \Sigma _{kk}}\log f(\mathbf{y}_{i\cdot })=-\frac{1%
}{2}\left( \mathbf{\Sigma }^{-1}\right) _{kk}+\frac{1}{2}\left( \mathbf{%
\Sigma }^{-1}\right) _{\cdot k}^{T}\mathrm{E}\left( \mathbf{\eta }_{i}%
\mathbf{\eta }_{i}^{T}|\mathbf{y}_{i\cdot }\right) \left( \mathbf{\Sigma }%
^{-1}\right) _{\cdot k}
\end{equation*}%
and 
\begin{equation*}
\frac{\partial }{\partial \Omega _{kk}}\log f(\mathbf{y}_{i\cdot })=-\frac{J%
}{2}\left( \mathbf{\Omega }^{-1}\right) _{kk}+\frac{1}{2}\left( \mathbf{%
\Omega }^{-1}\right) _{\cdot k}^{T}\sum_{j=1}^{J}\mathrm{E}\left( \mathbf{%
\xi }_{ij}\mathbf{\xi }_{ij}^{T}|\mathbf{y}_{i\cdot }\right) \left( \mathbf{%
\Omega }^{-1}\right) _{\cdot k},
\end{equation*}%
where $\left( \mathbf{\Sigma }^{-1}\right) _{\cdot k}$ and $\left( \mathbf{%
\Omega }^{-1}\right) _{\cdot k}$ denote the $k$th columns of $\mathbf{\Sigma 
}^{-1}$ and $\mathbf{\Omega }^{-1}$, respectively. Then, if we define $%
\widehat{\mathbf{\eta }_{i}^{\otimes 2}}=\mathrm{E}\left( \mathbf{\eta }%
_{i}\otimes \mathbf{\eta }_{i}|\mathbf{y}_{i\cdot }\right) $ and $\widehat{%
\mathbf{\xi }_{ij}^{\otimes 2}}=\mathrm{E}\left( \mathbf{\xi }_{ij}\otimes 
\mathbf{\xi }_{ij}|\mathbf{y}_{i\cdot }\right) $, after some algebra we
obtain: 
\begin{eqnarray*}
G_{kl} &=&-\frac{1}{4}\left( \mathbf{\Sigma }^{-1}\right) _{kk}\left( 
\mathbf{\Sigma }^{-1}\right) _{ll} \\
&&+\frac{1}{4}\left\{ \left( \mathbf{\Sigma }^{-1}\right) _{\cdot
k}^{T}\otimes \left( \mathbf{\Sigma }^{-1}\right) _{\cdot k}^{T}\right\} 
\mathrm{E}\Bigl(\widehat{\mathbf{\eta }_{i}^{\otimes 2}}\widehat{\mathbf{%
\eta }_{i}^{\otimes 2}}^{T}\Bigr)\left\{ \left( \mathbf{\Sigma }^{-1}\right)
_{\cdot l}\otimes \left( \mathbf{\Sigma }^{-1}\right) _{\cdot l}\right\} ,
\end{eqnarray*}%
\begin{eqnarray*}
G_{k,r+l} &=&-\frac{J}{4}\left( \mathbf{\Sigma }^{-1}\right) _{kk}\left( 
\mathbf{\Omega }^{-1}\right) _{ll} \\
&&+\frac{1}{4}\left\{ \left( \mathbf{\Sigma }^{-1}\right) _{\cdot
k}^{T}\otimes \left( \mathbf{\Sigma }^{-1}\right) _{\cdot k}^{T}\right\} 
\mathrm{E}\Bigl(\widehat{\mathbf{\eta }_{i}^{\otimes 2}}\sum_{j=1}^{J}%
\widehat{\mathbf{\xi }_{ij}^{\otimes 2}}^{T}\Bigr)\left\{ \left( \mathbf{%
\Omega }^{-1}\right) _{\cdot l}\otimes \left( \mathbf{\Omega }^{-1}\right)
_{\cdot l}\right\} ,
\end{eqnarray*}%
and 
\begin{eqnarray*}
G_{r+k,r+l} &=&-\frac{J^{2}}{4}\left( \mathbf{\Omega }^{-1}\right)
_{kk}\left( \mathbf{\Omega }^{-1}\right) _{ll} \\
&&+\frac{1}{4}\left\{ \left( \mathbf{\Omega }^{-1}\right) _{\cdot
k}^{T}\otimes \left( \mathbf{\Omega }^{-1}\right) _{\cdot k}^{T}\right\} 
\mathrm{E}\Bigl(\sum_{j=1}^{J}\widehat{\mathbf{\xi }_{ij}^{\otimes 2}}%
\sum_{j=1}^{J}\widehat{\mathbf{\xi }_{ij}^{\otimes 2}}^{T}\Bigr)\left\{
\left( \mathbf{\Omega }^{-1}\right) _{\cdot l}\otimes \left( \mathbf{\Omega }%
^{-1}\right) _{\cdot l}\right\} ,
\end{eqnarray*}%
for $k=1,\ldots ,r$ and $l=1,\ldots ,r$. As before, $\mathbf{\hat{G}}$ is
obtained replacing expectations by averages. Since the random-effect
estimators $\widehat{u_{ik}^{2}}$, $\widehat{v_{ijk}^{2}}$, $\widehat{%
\mathbf{\eta }_{i}^{\otimes 2}}$ and $\widehat{\mathbf{\xi }_{ij}^{\otimes 2}%
}$ are by-products of the EM algorithm, no extra computational costs are
incurred in computing $\mathbf{\hat{F}}$ and $\mathbf{\hat{G}}$.

Finally, we note that since $\hat{h}_{z}$ and $\hat{h}_{w}$ live in the
interval $[0,1]$, a transformation like $\arcsin \sqrt{h}$ usually provides
a better Normal approximation when $\hat{h}_{z}$ or $\hat{h}_{w}$ are close
to the boundaries. The asymptotic variance of $\arcsin \sqrt{\hat{h}}$ is
given by $\mathrm{avar}(\hat{h})/\{4\hat{h}(1-\hat{h})\}$. The simplest
procedure to derive a confidence interval for $\hat{h}$ in that case is to
construct a standard confidence interval for $\arcsin \sqrt{\hat{h}}$ and
then back-transform the endpoints.

\section{\label{sec:Simulations}Simulations}

In this section we study the finite-sample behavior of the new estimators by
simulation. The main goals are to determine if the new method \emph{(i)}
represents a substantial improvement over common functional ANOVA in
presence of time variability, \emph{(ii)} is at least comparable to the
naive approach of pre-warping the data using an existing warping method, 
\emph{(iii)} is robust to mild departures from the normality assumptions,
and \emph{(iv)} does not overfit, i.e.~is not worse than common functional
ANOVA in absence of time variability.

To this end we generated data from ten different models, all balanced, with $%
I=10$ groups and $J=5$ observations per group. The raw data (\ref%
{eq:raw_data_model}) was sampled on an equally-spaced time grid of $\nu =20$
points in $[0,1]$, and the noise variance was $\sigma ^{2}=.1^{2}$ in all
cases. The mean function was $\mu (t)=.6\varphi (t,.3,.1)+.4\varphi (t,.6,.1)
$ in all cases, where $\varphi (t,a,b)$ denotes the $N(a,b^{2})$ density
function. The models considered were the following:

\begin{enumerate}
\item One-component models (\ref{eq:KL_alpha}) and (\ref{eq:KL_beta}) with
no warping and $\phi _{1}(t)=\psi _{1}(t)=\varphi (t,.3,.1)/1.68$. The
variances were $\gamma _{1}=.2^{2}$ and $\lambda _{1}=.1^{2}$, so $h_{z}=.80$%
.

\item One-component models (\ref{eq:KL_alpha}) and (\ref{eq:KL_beta}) with
no warping but with different components for $\alpha (t)$ and $\beta (t)$: $%
\phi _{1}(t)$ as in Model 1, but $\psi _{1}(t)=\varphi (t,.6,.1)/1.68$. The
variances $\gamma _{1}$ and $\lambda _{1}$ were as in Model 1.

\item Same $\alpha (t)$ and $\beta (t)$ as in Model 1, but with a
Hermite-spline warping process $w(t)$ with knot $\tau _{0}=.3$ and variances 
$\Sigma =.2^{2}$ and $\Omega =.1^{2}$, so $h_{w}=.80$.

\item Same $\alpha (t)$ and $\beta (t)$ as in Model 2, with warping $w(t)$
as in Model 3.

\item Same $\alpha (t)$ and $\beta (t)$ as in Model 1, but with a warping
process $w(t)$ with knots $\mathbf{\tau }_{0}=(.3,.6)$ and covariance
matrices $\mathbf{\Sigma }=.2^{2}\mathbf{I}_{2}$ and $\mathbf{\Omega }=.1^{2}%
\mathbf{I}_{2}$, so $h_{w}=.80$ as before.

\item Same $\alpha (t)$ and $\beta (t)$ as in Model 2, with warping $w(t)$
as in Model 5.

\item Same as Model 4, but the random factors $U$ and $V$ in (\ref%
{eq:KL_alpha}) and (\ref{eq:KL_beta}) have Student's $t$ distributions with
4 degrees of freedom and scale parameters $\gamma _{1}^{1/2}=.2$ and $%
\lambda _{1}^{1/2}=.1$ (so the variance ratio is still $h_{z}=.80$).

\item Same as Model 4, but the random factors $U$ and $V$ in (\ref%
{eq:KL_alpha}) and (\ref{eq:KL_beta}) have contaminated Normal distributions 
$(1-\varepsilon )N(0,\gamma _{1})+\varepsilon N(0,k\gamma _{1})$ and $%
(1-\varepsilon )N(0,\lambda _{1})+\varepsilon N(0,k\lambda _{1})$
respectively, with $\varepsilon =.10$ and $k=5$ (the variance ratio is still 
$h_{z}=.80$).

\item Two-component models (\ref{eq:KL_alpha}) and (\ref{eq:KL_beta}), with $%
\phi _{1}(t)=\psi _{1}(t)=\varphi (t,.3,.1)/1.68$, $\phi _{2}(t)=\psi
_{2}(t)=(\varphi (t,.6,.1)/1.68-.105\phi _{1}(t))/.99$ (so that each pc is
associated with amplitude variation at each peak), variances $\gamma
_{1}=.2^{2}$, $\gamma _{2}=.1^{2}$, $\lambda _{1}=.1^{2}$, $\lambda
_{2}=.05^{2}$ (so $h_{z}=.80$ as in previous models), and a one-knot warping
process as in Model 3.

\item Same $\alpha (t)$ and $\beta (t)$ as in Model 9, with two-knot warping
as in Model 5.
\end{enumerate}

For each sample we computed the common (un-warped) ANOVA estimator, the
warped ANOVA estimator proposed in this paper, and a naive two-step warped
ANOVA estimator. The latter is computed as follows: first the curves are
aligned by least-squares registration (i.e.~minimizing $\sum_{i=1}^{n}\Vert
x_{i}\circ w_{i}-\mu \Vert ^{2}$ over $w_{i}$s in $\mathcal{W}_{\mathbf{\tau 
}_{0}}$) and then the common ANOVA estimators are computed on the warped
data. We used cubic B-splines with 10 equispaced knots as basis functions
for the functional parameters. As warping families we used interpolating
Hermite splines with $\tau _{0}=.3$ for models 1--4 and 7--9, and $\tau
_{0}=(.3,.6)$ for models 5--6 and 10. As error measures we used the bias,
the standard deviation and the root mean squared error, defined as follows:
if $f_{0}\in \mathcal{L}^{2}(I)$ and $\hat{f}$ is the estimator, then $%
\mathrm{bias}(\hat{f})=[\int \{\mathrm{E}\hat{f}(t)-f_{0}(t)\}^{2}dt]^{1/2}$%
, $\mathrm{sd}(\hat{f})=[\int \mathrm{E}\{\hat{f}(t)-\mathrm{E}\hat{f}%
(t)\}^{2}dt]^{1/2}$ and $\mathrm{rmse}(\hat{f})=\{\mathrm{bias}^{2}(\hat{f})+%
\mathrm{sd}^{2}(\hat{f})\}^{1/2}$. Some care must be taken with the
principal components, because their sign is undefined: to determine the
\textquotedblleft right\textquotedblright\ sign, we multiplied $\hat{\phi}%
_{1}$ and $\hat{\psi}_{1}$ by $\langle \hat{\phi}_{1},\phi _{1}\rangle $ and 
$\langle \hat{\psi}_{1},\psi _{1}\rangle $, respectively.

The estimation errors based on 200 Monte Carlo replications for each model
are shown in Table \ref{tab:simulations}. The effect of warping is more
clearly seen in the estimators of the principal components $\phi $ and $\psi 
$. The common ANOVA estimators, as expected, have the largest biases;
lacking a specific mechanism to handle time variability, common ANOVA
estimators $\hat{\phi}$ and $\hat{\psi}$ attempt to fit amplitude and phase
variability at the same time and get severely distorted compared to the true 
$\phi $ and $\psi $. The two warped estimators, on the other hand, can
handle phase variability well. The maximum-likelihood estimator proposed in
this paper always has smaller bias than the naive two-step approach; this is
to be expected, since the warping step of the two-step estimator minimizes
variation about the mean $\mu $ without taking into account amplitude
variability or the dependence structure in the data, whereas the maximum
likelihood estimator explicitly models $\phi $ and $\psi $. The down side of
the new estimators is that, as always, the bias reduction provided by the
more complex model is accompanied by a higher variance. However, looking at
the total root mean squared errors, we see that the new estimators
outperform the naive two-step estimators in almost all cases. This is also
true for the non-normal models 7 and 8, so the warped maximum likelihood
estimators are robust to mild departures from normality.

\section{\label{sec:Example}Example: beetle growth data}

In this section we study mass growth curves of flour beetles from birth to
pupation, from Irwin and Carter (2013). A total of 122 insects are
considered. This is a subset of a larger dataset that includes both siblings
and half-siblings, but in order to apply the one-way ANOVA model, which
assumes independence between groups, we consider only the half-siblings.
(The full data set can be modeled as a nested two-way ANOVA, with the mother
factor nested within the father factor.) The insects were sired by 29
different fathers, which will constitute the grouping variable. The number
of insects per father varies between 2 and 5, with a median of 4.

Part of the raw data is shown in Figure \ref{fig:curves}(a); for better
visualization we only plotted half of the sample curves. The mass measures
were taken about every 3 days early in the growth curve, and up to once per
day late in the growth curve. However, only 18 of the 122 larvae were
measured for mass for the first time on the day they hatched; 76 were
measured for mass for the first time on the second day, 22 on the third day,
5 on the fourth day, and one was not measured for mass until the seventh
day. Therefore, the starting points of the curves are unequal. The endpoints
are also irregular, because larvae reached pupation at different points
between days 16 and 25. However, while the unequal starting points are due
to missing data, the unequal endpoints are due to a well-defined biological
landmark which is reached at different times. Therefore we rescaled the time
grids so that all trajectories end at the median pupation day 19, but we did
not align the starting points at day 1. We also took logarithms to stabilize
the error variance. The rescaled log-data is shown in Figure \ref{fig:curves}%
(b).

These curves have a noticeable inflection point around day 15. This is
because in response to hormonal changes occurring prior to pupation, larvae
stop eating and start wandering in search of a place to pupate, and so lose
body mass. Therefore we fitted warped ANOVA models with a single warping
knot at $\tau _{0}=15$. As spline basis we used cubic B-splines with 7
equispaced knots; this gives a total of 9 basis functions, providing enough
flexibility without excessive irregularity. We considered several ANOVA
models with equal number of components for the main factor and the residual
term, ranging from 0 (mean-only model) to 3 components. The resulting
parameter estimators were:

\begin{itemize}
\item For $p=q=0$ (mean-only model): $\hat{\Sigma}=.013$, $\hat{\Omega}=.031$%
, $\hat{\sigma}=.181$.

\item For $p=q=1$: $\hat{\Sigma}=.010$, $\hat{\Omega}=.035$, $\hat{\gamma}%
=.323$, $\hat{\lambda}=.128$, $\hat{\sigma}=.138$.

\item For $p=q=2$: $\hat{\Sigma}=.010$, $\hat{\Omega}=.051$, $\mathbf{\hat{%
\gamma}}=(.344,.021)$, $\mathbf{\hat{\lambda}}=(.168,.010)$, $\hat{\sigma}%
=.124$.

\item For $p=q=3$: $\hat{\Sigma}=.005$, $\hat{\Omega}=.035$, $\mathbf{\hat{%
\gamma}}=(.426,.022,.005)$, $\mathbf{\hat{\lambda}}=(.186,.028,.012)$, $\hat{%
\sigma}=.121$.
\end{itemize}

Overall, it seems that a single principal component is sufficient to explain
amplitude variability, so we chose the model with $p=q=1$. The fitted curves 
$\hat{x}_{ij}(t)$ are shown in Figure \ref{fig:fits}(a) and we see that they
provide a good approximation to the data in Figure \ref{fig:curves}(b). The
estimated warping functions $\hat{w}_{ij}(t)$ are shown in Figure \ref%
{fig:fits}(b); the time variability around day 15, which is substantial, is
captured well by these curves. The amplitude principal components $\hat{\phi}%
(t)$ and $\hat{\psi}(t)$ are shown in Figure \ref{fig:fits}(c); to
facilitate interpretation of the components we plotted $\hat{\mu}(t)$
together with $\hat{\mu}(t)\pm \hat{\phi}(t)$ in Figure \ref{fig:fits}(d).
We see that $\hat{\phi}(t)$ and $\hat{\psi}(t)$, which are very similar,
explain variation in overall mass: individuals with positive pc scores tend
to have trajectories above the mean and individuals with negative pc scores
tend to have trajectories below the mean.

The similarity between $\hat{\phi}(t)$ and $\hat{\psi}(t)$ has a biological
explanation: the main factor of the ANOVA model represents the genetic
contribution of the father, while the residual term represents the genetic
contribution of the mother together with environmental factors (see
e.g.~Heckman, 2003, sec.~3). For a population in Hardy-Weinberg equilibrium,
the genetic contribution of both parents is identical, so the $\hat{\phi}%
_{k}(t)$s and the $\hat{\psi}_{k}(t)$s will be similar if the environmental
factors are not very strong. Supporting this result is the fact that Irwin
and Carter (2013) showed that most of the phenotypic variance was explained
by genetic effects in most parts of the growth curve.

The amplitude principal components reveal a very interesting biological
result that was not apparent in the original analysis of the raw data in
Irwin and Carter (2013): very little variation in amplitude exists at the
inflection point at day 15 (Figure \ref{fig:fits}(c) and \ref{fig:fits}(d)).
This indicates that the beetles have a target peak mass that is reached
prior to entry into the wandering phase, which suggests that the target peak
mass must be reached before pupation can begin, and that selection for that
peak mass (or a related physiological trait) may occur. Interestingly the
warping functions shown in Figure \ref{fig:fits}(b), as well as the original
analysis in Irwin and Carter (2013) demonstrate there is substantial
variation at the age at which peak mass is reached. In combination these two
results provide a basis for future experiments investigating physiological
mechanisms, genetic underpinnings and evolutionary implications of size and
age of peak mass.

The variance ratios for the amplitude and warping components are $\hat{h}%
_{z}=.72$ and $\hat{h}_{w}=.23$, with respective asymptotic standard
deviations $.15$ and $.13$. The bootstrap distributions of $\hat{h}_{z}$ and 
$\hat{h}_{w}$ are shown in Web Appendix D; the bootstrap standard deviations
are $.20$ and $.16$ respectively, not far from the asymptotic values, but
the Normal approximation is more accurate for the transformations $\arcsin 
\sqrt{\hat{h}_{z}}$ and $\arcsin \sqrt{\hat{h}_{w}}$. The 90\% asymptotic
confidence intervals obtained by the back-transformation method are $%
(.45,.92)$ for $h_{z}$ and $(.06,.47)$ for $h_{w}$. Clearly the father
effect is strong on the amplitude component, but weak on the warping
component. This can be cross-checked by applying the classical ANOVA $F$%
-test on the estimated random effects $\hat{\theta}_{ij}$s: it yields a $p$%
-value of $0.058$ for the hypothesis of no father effect (the reasonableness
of the normality assumption on the random factors is also discussed in Web
Appendix D.) The reason the father effect is weak on the warping component
is that we removed a lot of time variability by aligning the endpoints at
the median pupation day. In fact, the ANOVA $F$-test on the original
endpoints yields a $p$-value of $0.020$, indicating that there is a
significant father\ effect on the date of pupation; this is also supported
by Irwin and Carter (2013) demonstrating a highly significant heritability
(genetic variance ratio) for date of pupation in the full sample. But once
the endpoints are aligned, the time variability that remains, although still
substantial, does not have a strong father effect.

If we assume $\phi =\psi $, which is not unreasonable given Figure \ref%
{fig:fits}(c), then (\ref{eq:z_model}) comes down to $z_{ij}(t)=\mu
(t)+(U_{i}+V_{ij})\phi (t)$ and the classical ANOVA $F$-test can be applied
to $\{\hat{U}_{i}+\hat{V}_{ij}\}$. This gives a very significant $F$-value $%
11.03$ with $p$-value $0.00$, confirming that the father effect is very
strong on the amplitude variability of the growth curves.

As indicated at the end of Section \ref{sec:Model}, a more general model
with correlations between amplitude and warping components can be set up. We
fitted a one-component model with correlations for these data and obtained
estimators $\hat{\phi}_{1}$ and $\hat{\psi}_{1}$ very similar to the ones
obtained above, and correlations $\widehat{\mathrm{corr}}(U_{i},\mathbf{\eta 
}_{i})=.18$ and $\widehat{\mathrm{corr}}(V_{ij},\mathbf{\xi }_{ij})=.17$,
which do not seem very significant. The statistical significance of these
correlations could be studied, for instance, by bootstrap confidence
intervals, but for brevity's sake we will not do it here.

Finally, it is important to note that for the unaligned raw data, variation
in the length of the larval period and the onset of the wandering phase
resulted in crossing of family curves late in the larval period (Irwin and
Carter, 2013). After application of the warping method, the warped curves
are aligned by peak body mass at the onset of the wandering phase, resulting
in family curves late in the larval period that maintain relative positions
similar to early in the larval period. This realignment undoubtedly will
facilitate estimation of genetic components of variance, a proposition that
we can test in the future.

\section{Supplementary materials}

Web Appendices referenced in Sections 3--6 and Matlab programs implementing
the new estimators are available with this paper at the Biometrics website
on Wiley Online Library.

\section*{Acknowledgements}

This research was supported by National Science Foundation grant DMS-1006281
to Daniel Gervini, NSF grant EF-0328594 to Patrick A.~Carter and a grant
from the National Institute for Mathematical and Biological Synthesis to
Patrick A.~Carter.

\section*{References}

\begin{description}
\item Ash, R.B., and Gardner, M.F. (1975). \emph{Topics in Stochastic
Processes}. Academic Press.

\item Bigot, J., and Gadat, S. (2010). A deconvolution approach to
estimation of a common shape in a shifted curves model. \emph{The Annals of
Statistics} \textbf{38} 2422--2464.

\item Bookstein, F. L. (1997). \emph{Morphometric tools for landmark data:
geometry and biology}. Cambridge University Press.

\item Brumback, L.C., and Lindstrom, M. (2004). Self modeling with flexible,
random time transformations. \emph{Biometrics} \textbf{60} 461--470.

\item Chen, H., and Wang, Y. (2011). A penalized spline approach to
functional mixed effects model analysis. \emph{Biometrics} \textbf{67}
861--870.

\item Claeskens, G., Silverman, B. W., and Slaets, L. (2010). A
multiresolution approach to time warping achieved by a Bayesian
prior--posterior transfer fitting strategy. \emph{Journal of the Royal
Statistical Society Series B} \textbf{72} 673--694.

\item Di, C. Z., Crainiceanu, C. M., Caffo, B. S., and Punjabi, N. M.
(2009). Multilevel functional principal component analysis. \emph{The Annals
of Applied Statistics} \textbf{3} 458--488.

\item Fritsch, F. N., and Carlson, R. E. (1980). Monotone piecewise cubic
interpolation. \emph{SIAM Journal of Numerical Analysis} \textbf{17}
238--246.

\item Gervini, D. (2009). Detecting and handling outlying trajectories in
irregularly sampled functional datasets. \emph{The Annals of Applied
Statistics} \textbf{3} 1758--1775.

\item Gervini, D., and Gasser, T. (2004). Self-modelling warping functions. 
\emph{Journal of the Royal Statistical Society Series B} \textbf{66}
959--971.

\item Gervini, D., and Gasser, T. (2005). Nonparametric maximum likelihood
estimation of the structural mean of a sample of curves. \emph{Biometrika} 
\textbf{92} 801--820.

\item Gohberg, I., Goldberg, S., and Kaashoek, M. A. (2003). \emph{Basic
Classes of Linear Operators}. Basel: Birkh\"{a}user Verlag.

\item Gomulkiewicz, R., and Beder, J. H. (1996). The selection gradient of
an infinite-dimensional trait. \emph{SIAM Journal of Applied Mathematics} 
\textbf{56} 509--523.

\item Guo, W. (2002). Functional mixed effects models. \emph{Biometrics} 
\textbf{58} 121--128.

\item Heckman, N. E. (2003). Functional data analysis in evolutionary
biology. In \emph{Recent Advances and Trends in Nonparametric Statistics.}
Elsevier.

\item Huey, R. B., and Kingsolver, J. G. (1989). Evolution of thermal
sensitivity of ectotherm performance. \emph{Trends in Ecology and Evolution} 
\textbf{4} 131--135.

\item Irwin, K.K., and Carter, P.A. (2013). Constraints on the evolution of
function-valued traits: a study of growth in \emph{Tribolium casteneum}. 
\emph{Journal of Evolutionary Biology }(in press).

\item Izem, R., and Kingsolver, J. G. (2005). Variation in continuous
reaction norms: quantifying directions of biological interest. \emph{The
American Naturalist}, \textbf{166}, 277--289.

\item Jupp, D. L. B. (1978) Approximation to data by splines with free
knots. \emph{SIAM Journal of Numerical Analysis} \textbf{15} 328--343.

\item Kingsolver, J. G., Gomulkiewicz, R., and Carter, P. A. (2002).
Variation, selection and evolution of function-valued traits. In \emph{%
Microevolution Rate, Pattern, Process,} pp.~87--104. Springer.

\item Kirkpatrick, M., and Heckman, N. (1989). A quantitative genetic model
for growth, shape, reaction norms, and other infinite-dimensional
characters. \emph{Journal of Mathematical Biology} \textbf{27} 429--450.

\item Kneip, A., and Engel, J. (1995). Model estimation in nonlinear
regression under shape invariance. \emph{The Annals of Statistics} \textbf{23%
} 551--570.

\item Kneip, A., Li, X., MacGibbon, K. B., and Ramsay, J. O. (2000). Curve
registration by local regression. \emph{Canadian Journal of Statistics} 
\textbf{28} 19--29.

\item Kneip, A., and Ramsay, J. O. (2008). Combining registration and
fitting for functional models. \emph{Journal of the American Statistical
Association} \textbf{103} 1155--1165.

\item Manikkam, M., Guerrero-Bosagna, C., Tracey, R., Haque, M., and
Skinner, M. (2012). Transgenerational actions of environmental compounds on
reproductive disease and identification of epigenetic biomarkers of
ancestral exposures. \emph{PLoS One 7}.

\item Meyer, K., and Kirkpatrick, M. (2005). Up hill, down dale:
quantitative genetics of curvaceous traits. \emph{Philosophical Transactions
of the Royal Society B: Biological Sciences}, \textbf{360}, 1443--1455.

\item Morris, J. S., and Carroll, R. J. (2006). Wavelet-based functional
mixed models. \emph{Journal of the Royal Statistical Society Series B} 
\textbf{68} 179--199.

\item M\"{u}ller, H. G. (2008). Functional modeling of longitudinal data. In 
\emph{Longitudinal data analysis. Handbooks of modern statistical methods}.
Chapman \& Hall/CRC, New York, pp.~223--252.

\item Ragland, G.J., and Carter, P.A. (2004). Genetic constraints on the
evolution of growth and life history traits in the salamander Ambystoma
macrodactylum. \emph{Heredity} \textbf{92 }569--578.

\item Ramsay, J. O., and Li, X. (1998). Curve registration. \emph{Journal of
the Royal Statistical Society Series B} \textbf{60} 351--363.

\item Rice, J. A. (2004). Functional and longitudinal data analysis:
Perspectives on smoothing. \emph{Statistica Sinica} \textbf{14} 631--648.

\item Sangalli, L.M., Secchi, P., Vantini, S., and Vitelli, V. (2010).
k-mean alignment for curve clustering. \emph{Computational Statistics and
Data Analysis} \textbf{54} 1219--1233.

\item Skinner, M. K., Manikkam, M., and Guerrero-Bosagna, C. (2010).
Epigenetic transgenerational actions of environmental factors in disease
etiology. \emph{Trends in Endocrinology and Metabolism} \textbf{21} 214--222.

\item Tang, R., and M\"{u}ller, H. G. (2008). Pairwise curve synchronization
for functional data. \emph{Biometrika} \textbf{95} 875--889.

\item Telesca, D., and Inoue, L. Y. (2008). Bayesian hierarchical curve
registration. \emph{Journal of the American Statistical Association} \textbf{%
103} 328--339.

\item Wang, K. and Gasser, T. (1999). Synchronizing sample curves
nonparametrically. \emph{The Annals of Statistics} \textbf{27} 439--460.
\end{description}

\newpage

\bigskip 
\begin{table}[h] \centering%
\begin{sideways}%
\begin{tabular}{lccccccllllccccccccc}
& \multicolumn{9}{c}{\small Model 1} &  & \multicolumn{9}{c}{\small Model 2}
\\ 
& \multicolumn{3}{c}{\small bias} & \multicolumn{3}{c}{\small sd} & 
\multicolumn{3}{c}{\small rmse} &  & \multicolumn{3}{c}{\small bias} & 
\multicolumn{3}{c}{\small sd} & \multicolumn{3}{c}{\small rmse} \\ 
& {\small C} & {\small 2s} & {\small ML} & {\small C} & {\small 2s} & 
{\small ML} & {\small C} & {\small 2s} & {\small ML} &  & {\small C} & 
{\small 2s} & {\small ML} & {\small C} & {\small 2s} & {\small ML} & {\small %
C} & {\small 2s} & {\small ML} \\ 
${\small \hat{\mu}}$ & \multicolumn{1}{l}{\small .008} & {\small .016} & 
\multicolumn{1}{l}{\small .008} & \multicolumn{1}{|l}{\small .055} & {\small %
.055} & \multicolumn{1}{l}{\small .055} & \multicolumn{1}{|l}{\small .055} & 
{\small .057} & {\small .056} &  & \multicolumn{1}{l}{\small .008} & {\small %
.017} & \multicolumn{1}{l}{\small .008} & \multicolumn{1}{|l}{\small .051} & 
{\small .052} & \multicolumn{1}{l}{\small .057} & \multicolumn{1}{|c}{\small %
.051} & {\small .054} & {\small .058} \\ 
${\small \hat{\phi}}_{1}$ & \multicolumn{1}{l}{\small .007} & {\small .127}
& \multicolumn{1}{l}{\small .048} & \multicolumn{1}{|l}{\small .064} & 
{\small .067} & \multicolumn{1}{l}{\small .069} & \multicolumn{1}{|l}{\small %
.064} & {\small .143} & {\small .084} &  & \multicolumn{1}{l}{\small .172} & 
{\small .219} & \multicolumn{1}{l}{\small .130} & \multicolumn{1}{|l}{\small %
.078} & {\small .077} & \multicolumn{1}{l}{\small .101} & \multicolumn{1}{|c}%
{\small .189} & {\small .233} & {\small .165} \\ 
${\small \hat{\psi}}_{1}$ & \multicolumn{1}{l}{\small .014} & {\small .127}
& \multicolumn{1}{l}{\small .032} & \multicolumn{1}{|l}{\small .120} & 
{\small .121} & \multicolumn{1}{l}{\small .116} & \multicolumn{1}{|l}{\small %
.121} & {\small .175} & {\small .121} &  & \multicolumn{1}{l}{\small .013} & 
{\small .184} & \multicolumn{1}{l}{\small .050} & \multicolumn{1}{|l}{\small %
.112} & {\small .111} & \multicolumn{1}{l}{\small .136} & \multicolumn{1}{|c}%
{\small .113} & {\small .215} & {\small .145} \\ 
& \multicolumn{1}{l}{} &  & \multicolumn{1}{l}{} & \multicolumn{1}{l}{} &  & 
\multicolumn{1}{l}{} &  &  &  &  & \multicolumn{1}{l}{} &  & 
\multicolumn{1}{l}{} & \multicolumn{1}{l}{} &  & \multicolumn{1}{l}{} &  & 
&  \\ 
& \multicolumn{9}{c}{\small Model 3} &  & \multicolumn{9}{c}{\small Model 4}
\\ 
${\small \hat{\mu}}$ & \multicolumn{1}{l}{\small .073} & {\small .033} & 
\multicolumn{1}{l}{\small .013} & \multicolumn{1}{|l}{\small .081} & {\small %
.083} & \multicolumn{1}{l}{\small .088} & \multicolumn{1}{|l}{\small .109} & 
{\small .089} & {\small .089} &  & \multicolumn{1}{r}{\small .071} & 
\multicolumn{1}{r}{\small .032} & \multicolumn{1}{r}{\small .014} & 
\multicolumn{1}{|r}{\small .085} & \multicolumn{1}{r}{\small .085} & 
\multicolumn{1}{r}{\small .093} & \multicolumn{1}{|r}{\small .111} & 
\multicolumn{1}{r}{\small .091} & \multicolumn{1}{r}{\small .094} \\ 
${\small \hat{\phi}}_{1}$ & \multicolumn{1}{l}{\small .175} & {\small .129}
& \multicolumn{1}{l}{\small .051} & \multicolumn{1}{|l}{\small .530} & 
{\small .109} & \multicolumn{1}{l}{\small .145} & \multicolumn{1}{|l}{\small %
.559} & {\small .168} & {\small .154} &  & \multicolumn{1}{r}{\small .109} & 
\multicolumn{1}{r}{\small .213} & \multicolumn{1}{r}{\small .087} & 
\multicolumn{1}{|r}{\small .422} & \multicolumn{1}{r}{\small .108} & 
\multicolumn{1}{r}{\small .173} & \multicolumn{1}{|r}{\small .436} & 
\multicolumn{1}{r}{\small .239} & \multicolumn{1}{r}{\small .194} \\ 
${\small \hat{\psi}}_{1}$ & \multicolumn{1}{l}{\small .327} & {\small .126}
& \multicolumn{1}{l}{\small .027} & \multicolumn{1}{|l}{\small .736} & 
{\small .145} & \multicolumn{1}{l}{\small .167} & \multicolumn{1}{|l}{\small %
.806} & {\small .192} & {\small .169} &  & \multicolumn{1}{r}{\small 1.095}
& \multicolumn{1}{r}{\small .192} & \multicolumn{1}{r}{\small .057} & 
\multicolumn{1}{|r}{\small .293} & \multicolumn{1}{r}{\small .135} & 
\multicolumn{1}{r}{\small .226} & \multicolumn{1}{|r}{\small 1.134} & 
\multicolumn{1}{r}{\small .235} & \multicolumn{1}{r}{\small .233} \\ 
& \multicolumn{1}{l}{} &  & \multicolumn{1}{l}{} & \multicolumn{1}{l}{} &  & 
\multicolumn{1}{l}{} &  &  &  &  & \multicolumn{1}{l}{} &  & 
\multicolumn{1}{l}{} & \multicolumn{1}{l}{} &  & \multicolumn{1}{l}{} &  & 
&  \\ 
& \multicolumn{9}{c}{\small Model 5} &  & \multicolumn{9}{c}{\small Model 6}
\\ 
${\small \hat{\mu}}$ & \multicolumn{1}{l}{\small .106} & {\small .041} & 
\multicolumn{1}{l}{\small .024} & \multicolumn{1}{|l}{\small .090} & {\small %
.094} & \multicolumn{1}{l}{\small .119} & \multicolumn{1}{|l}{\small .139} & 
{\small .103} & {\small .122} &  & \multicolumn{1}{r}{\small .104} & 
\multicolumn{1}{r}{\small .048} & \multicolumn{1}{r}{\small .023} & 
\multicolumn{1}{|r}{\small .088} & \multicolumn{1}{r}{\small .094} & 
\multicolumn{1}{r}{\small .113} & \multicolumn{1}{|r}{\small .136} & 
\multicolumn{1}{r}{\small .106} & \multicolumn{1}{r}{\small .116} \\ 
${\small \hat{\phi}}_{1}$ & \multicolumn{1}{l}{\small .226} & {\small .233}
& \multicolumn{1}{l}{\small .114} & \multicolumn{1}{|l}{\small .603} & 
{\small .135} & \multicolumn{1}{l}{\small .302} & \multicolumn{1}{|l}{\small %
.644} & {\small .269} & {\small .323} &  & \multicolumn{1}{r}{\small .179} & 
\multicolumn{1}{r}{\small .282} & \multicolumn{1}{r}{\small .047} & 
\multicolumn{1}{|r}{\small .521} & \multicolumn{1}{r}{\small .130} & 
\multicolumn{1}{r}{\small .245} & \multicolumn{1}{|r}{\small .551} & 
\multicolumn{1}{r}{\small .310} & \multicolumn{1}{r}{\small .250} \\ 
${\small \hat{\psi}}_{1}$ & \multicolumn{1}{l}{\small .468} & {\small .230}
& \multicolumn{1}{l}{\small .082} & \multicolumn{1}{|l}{\small .838} & 
{\small .163} & \multicolumn{1}{l}{\small .263} & \multicolumn{1}{|l}{\small %
.960} & {\small .282} & {\small .275} &  & \multicolumn{1}{r}{\small 1.013}
& \multicolumn{1}{r}{\small .380} & \multicolumn{1}{r}{\small .154} & 
\multicolumn{1}{|r}{\small .329} & \multicolumn{1}{r}{\small .281} & 
\multicolumn{1}{r}{\small .362} & \multicolumn{1}{|r}{\small 1.065} & 
\multicolumn{1}{r}{\small .473} & \multicolumn{1}{r}{\small .393} \\ 
&  &  &  &  &  &  &  &  &  &  &  &  &  &  &  &  &  &  &  \\ 
& \multicolumn{9}{c}{\small Model 7} &  & \multicolumn{9}{c}{\small Model 8}
\\ 
${\small \hat{\mu}}$ & {\small .070} & {\small .042} & {\small .020} & 
\multicolumn{1}{|c}{\small .093} & {\small .099} & {\small .129} & 
\multicolumn{1}{|l}{\small .116} & {\small .108} & {\small .130} &  & 
{\small .072} & {\small .036} & {\small .016} & \multicolumn{1}{|c}{\small %
.083} & {\small .086} & {\small .100} & \multicolumn{1}{|c}{\small .110} & 
{\small .093} & {\small .101} \\ 
${\small \hat{\phi}}_{1}$ & {\small .082} & {\small .244} & {\small .080} & 
\multicolumn{1}{|c}{\small .322} & {\small .122} & {\small .195} & 
\multicolumn{1}{|l}{\small .332} & {\small .273} & {\small .211} &  & 
{\small .132} & {\small .242} & {\small .096} & \multicolumn{1}{|c}{\small %
.385} & {\small .111} & {\small .182} & \multicolumn{1}{|c}{\small .407} & 
{\small .266} & {\small .206} \\ 
${\small \hat{\psi}}_{1}$ & {\small .874} & {\small .205} & {\small .043} & 
\multicolumn{1}{|c}{\small .389} & {\small .154} & {\small .221} & 
\multicolumn{1}{|l}{\small .957} & {\small .257} & {\small .225} &  & 
{\small 1.001} & {\small .199} & {\small .069} & \multicolumn{1}{|c}{\small %
.354} & {\small .126} & {\small .205} & \multicolumn{1}{|c}{\small 1.061} & 
{\small .235} & {\small .217} \\ 
&  &  &  &  &  &  &  &  &  &  &  &  &  &  &  &  &  &  &  \\ 
& \multicolumn{9}{c}{\small Model 9} &  & \multicolumn{9}{c}{\small Model 10}
\\ 
${\small \hat{\mu}}$ & \multicolumn{1}{r}{\small .072} & \multicolumn{1}{r}%
{\small .035} & \multicolumn{1}{r}{\small .019} & \multicolumn{1}{|r}{\small %
.086} & \multicolumn{1}{r}{\small .084} & \multicolumn{1}{r}{\small .105} & 
\multicolumn{1}{|r}{\small .112} & \multicolumn{1}{r}{\small .091} & 
\multicolumn{1}{r}{\small .106} &  & \multicolumn{1}{r}{\small .106} & 
\multicolumn{1}{r}{\small .059} & \multicolumn{1}{r}{\small .042} & 
\multicolumn{1}{|r}{\small .097} & \multicolumn{1}{r}{\small .100} & 
\multicolumn{1}{r}{\small .136} & \multicolumn{1}{|r}{\small .143} & 
\multicolumn{1}{r}{\small .116} & \multicolumn{1}{r}{\small .143} \\ 
${\small \hat{\phi}}_{1}$ & \multicolumn{1}{r}{\small .242} & 
\multicolumn{1}{r}{\small .206} & \multicolumn{1}{r}{\small .172} & 
\multicolumn{1}{|r}{\small .633} & \multicolumn{1}{r}{\small .337} & 
\multicolumn{1}{r}{\small .559} & \multicolumn{1}{|r}{\small .678} & 
\multicolumn{1}{r}{\small .395} & \multicolumn{1}{r}{\small .585} &  & 
\multicolumn{1}{r}{\small .346} & \multicolumn{1}{r}{\small .262} & 
\multicolumn{1}{r}{\small .236} & \multicolumn{1}{|r}{\small .663} & 
\multicolumn{1}{r}{\small .353} & \multicolumn{1}{r}{\small .637} & 
\multicolumn{1}{|r}{\small .748} & \multicolumn{1}{r}{\small .439} & 
\multicolumn{1}{r}{\small .679} \\ 
${\small \hat{\phi}}_{2}$ & \multicolumn{1}{r}{\small .357} & 
\multicolumn{1}{r}{\small .602} & \multicolumn{1}{r}{\small .418} & 
\multicolumn{1}{|r}{\small .571} & \multicolumn{1}{r}{\small .903} & 
\multicolumn{1}{r}{\small .801} & \multicolumn{1}{|r}{\small .673} & 
\multicolumn{1}{r}{\small 1.085} & \multicolumn{1}{r}{\small .904} &  & 
\multicolumn{1}{r}{\small .503} & \multicolumn{1}{r}{\small .708} & 
\multicolumn{1}{r}{\small .469} & \multicolumn{1}{|r}{\small .742} & 
\multicolumn{1}{r}{\small .825} & \multicolumn{1}{r}{\small .845} & 
\multicolumn{1}{|r}{\small .896} & \multicolumn{1}{r}{\small 1.087} & 
\multicolumn{1}{r}{\small .967} \\ 
${\small \hat{\psi}}_{1}$ & \multicolumn{1}{r}{\small .383} & 
\multicolumn{1}{r}{\small .258} & \multicolumn{1}{r}{\small .139} & 
\multicolumn{1}{|r}{\small .761} & \multicolumn{1}{r}{\small .209} & 
\multicolumn{1}{r}{\small .427} & \multicolumn{1}{|r}{\small .852} & 
\multicolumn{1}{r}{\small .332} & \multicolumn{1}{r}{\small .449} &  & 
\multicolumn{1}{r}{\small .521} & \multicolumn{1}{r}{\small .312} & 
\multicolumn{1}{r}{\small .156} & \multicolumn{1}{|r}{\small .835} & 
\multicolumn{1}{r}{\small .232} & \multicolumn{1}{r}{\small .510} & 
\multicolumn{1}{|r}{\small .984} & \multicolumn{1}{r}{\small .389} & 
\multicolumn{1}{r}{\small .533} \\ 
${\small \hat{\psi}}_{2}$ & \multicolumn{1}{r}{\small .839} & 
\multicolumn{1}{r}{\small .321} & \multicolumn{1}{r}{\small .387} & 
\multicolumn{1}{|r}{\small .888} & \multicolumn{1}{r}{\small .284} & 
\multicolumn{1}{r}{\small .771} & \multicolumn{1}{|r}{\small 1.222} & 
\multicolumn{1}{r}{\small .429} & \multicolumn{1}{r}{\small .863} &  & 
\multicolumn{1}{r}{\small .780} & \multicolumn{1}{r}{\small .552} & 
\multicolumn{1}{r}{\small .448} & \multicolumn{1}{|r}{\small .910} & 
\multicolumn{1}{r}{\small .471} & \multicolumn{1}{r}{\small .826} & 
\multicolumn{1}{|r}{\small 1.211} & \multicolumn{1}{r}{\small .726} & 
\multicolumn{1}{r}{\small .940}%
\end{tabular}%
\end{sideways}%
\caption{Simulation Results. Biase, standard deviation and root mean squared error for common (C),
two-step (2s) and maximum likelihood (ML) ANOVA estimators.}\label%
{tab:simulations}%
\end{table}%

\newpage

\FRAME{fhFU}{6.4636in}{2.2329in}{0pt}{\Qcb{Flour Beetle Growth Example. (a)
Raw mass trajectories; (b) log-trajectories re-scaled to common endpoint.}}{%
\Qlb{fig:curves}}{curves.eps}{\special{language "Scientific Word";type
"GRAPHIC";maintain-aspect-ratio TRUE;display "USEDEF";valid_file "F";width
6.4636in;height 2.2329in;depth 0pt;original-width 8.3385in;original-height
6.1609in;cropleft "0.0663";croptop "1";cropright "1";cropbottom "0";filename
'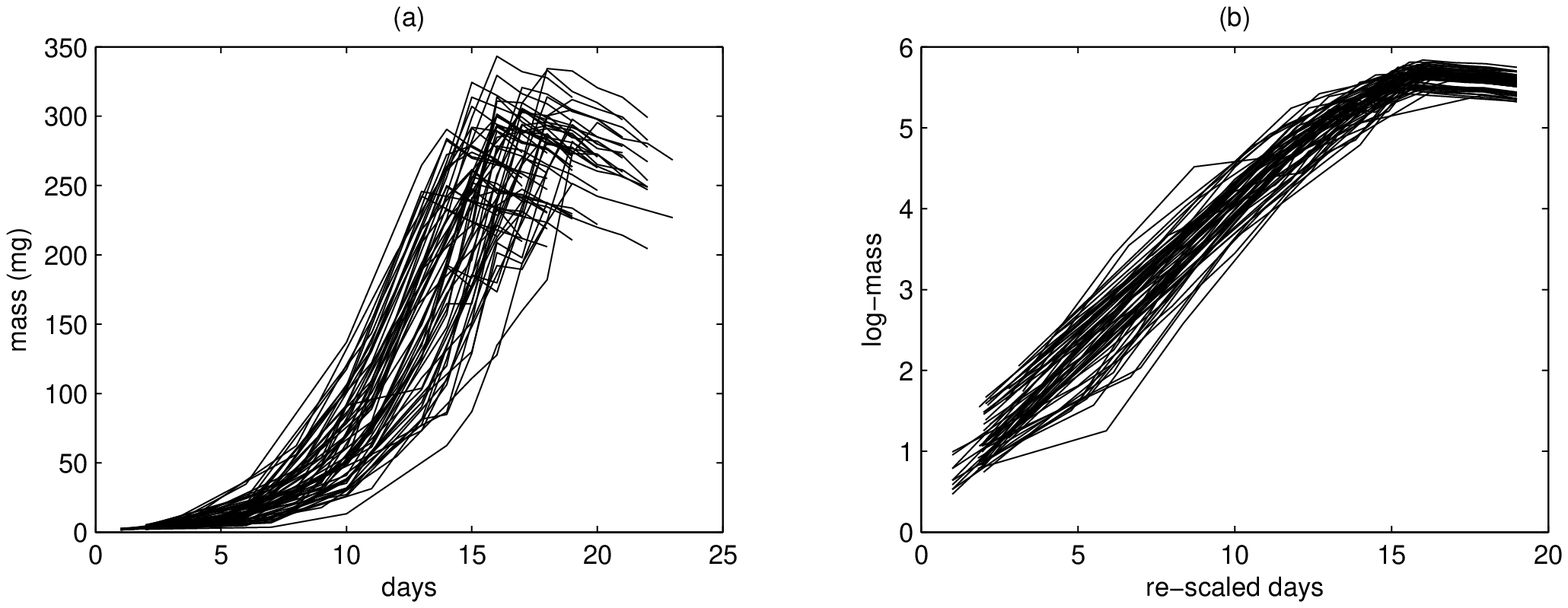';file-properties "XNPEU";}}

\newpage

\FRAME{fhFU}{6.4454in}{5.0246in}{0pt}{\Qcb{Flour Beetle Growth Example. (a)
Fitted trajectories using warped ANOVA model; (b) warping functions; (c)
principal component of the main factor, $\hat{\protect\phi}(t)$ (solid
line), and of the residual term, $\hat{\protect\psi}(t)$ (dashed line); (d)
estimated mean $\hat{\protect\mu}(t)$ (solid line), $\hat{\protect\mu}(t)+%
\hat{\protect\phi}(t)$ (dash-dot line), and $\hat{\protect\mu}(t)-\hat{%
\protect\phi}(t)$ (dotted line). }}{\Qlb{fig:fits}}{fits.eps}{\special%
{language "Scientific Word";type "GRAPHIC";maintain-aspect-ratio
TRUE;display "USEDEF";valid_file "F";width 6.4454in;height 5.0246in;depth
0pt;original-width 9.4537in;original-height 3.022in;cropleft
"0.0728";croptop "1";cropright "1";cropbottom "0";filename
'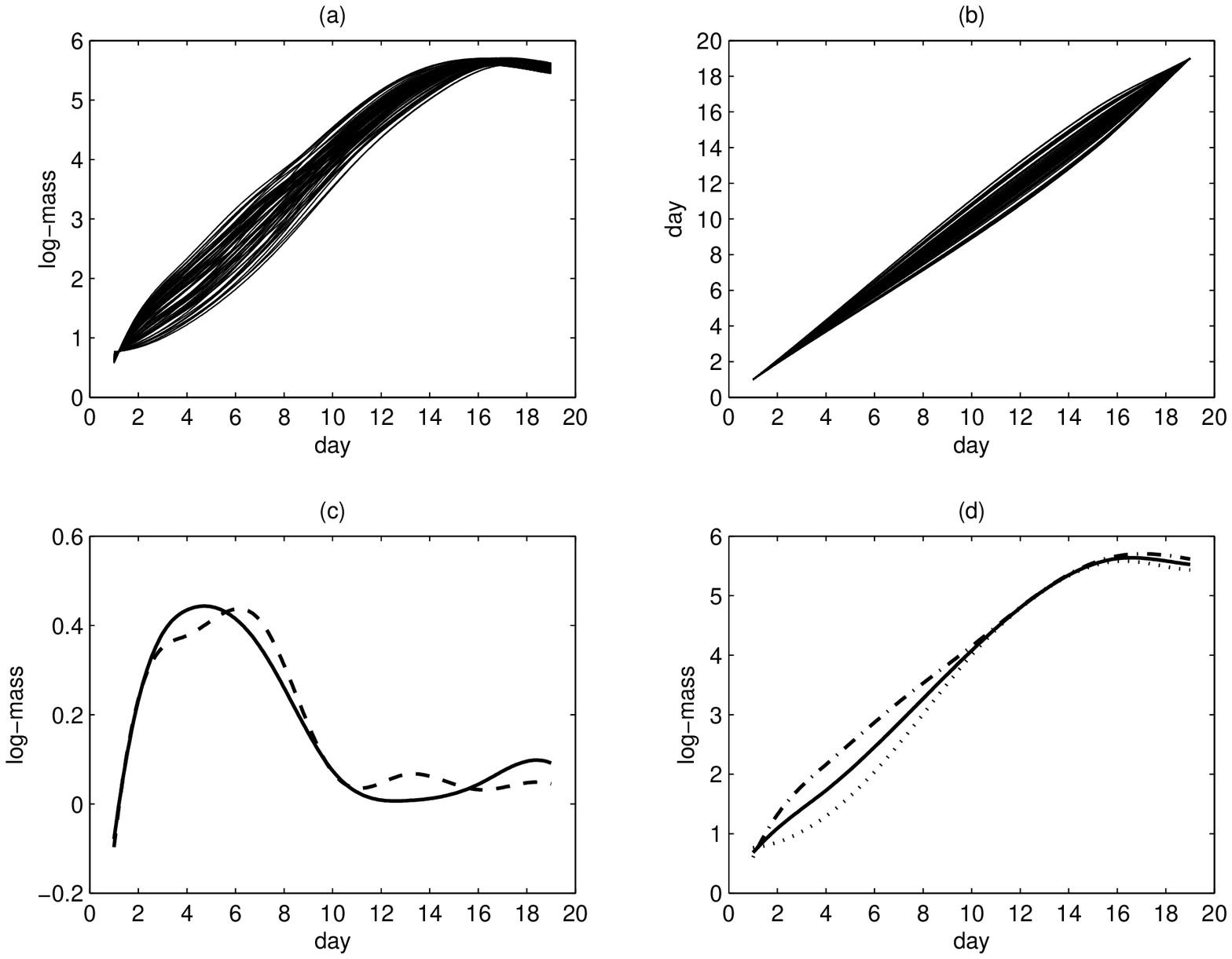';file-properties "XNPEU";}}

\end{document}